\documentclass[fleqn,10pt]{manuscript}
\usepackage{soul}
\usepackage{upgreek}

\title{Terahertz microresonators for material characterisation}

\author[1]{Dominik Walter Vogt}

\affil[1]{The Dodd-Walls Centre for Photonic and Quantum Technologies, Department of Physics, The University of Auckland, Auckland 1010, New Zealand}

\corrauthor[1]{Dominik Walter Vogt}{d.vogt@auckland.ac.nz}

\begin{abstract}
Terahertz (THz) technology is rapidly evolving, and the advancement of data and information processing devices is essential. Silicon THz microresonators provide perfect platforms to develop compact and integrated devices that could transform THz technology. Here we present a systematic study on the key figure of merit of silicon THz disc microresonators - the quality factor (Q-factor) - in dependence on the substrate's resistivity. Our results show that the Q-factor depends linearly on the resistivity and a variation in resistivity from 10\,k$\Omega$cm to 15\,k$\Omega$cm changes the Q-factor from 50k to 76k at 0.6\,THz. Moreover, we experimentally determine that the silicon material absorption is inversely proportional to the substrate's resistivity. In general, the presented methodology is ideally suited to precisely measure the material absorption of low-loss materials in the THz domain, which is challenging using conventional THz spectroscopy.  
\end{abstract}

\begin{document}

\flushbottom
\maketitle
\thispagestyle{empty}

\section*{Introduction}
Silicon THz microresonators continue to attract significant interest as fundamental building blocks for essential devices like sensors, filters, isolators, and temporal differentiators \cite{vogt2020terahertz, yuan2019mode, yuan2019tunable, gandhi2021microresonator,zhou2021photonics, wang2019voltage, yuan2021chip}. In particular, silicon THz microresonators stand out due to very high Q-factors of up to 120,000 as well as the convenient implementation pathway for integrated devices \cite{vogt2020subwavelength,wang2019chip}. Various designs ranging from sub-wavelength thin discs achieving record Q-factors and sensitivities to racetrack ring resonators ideal for on-chip integration have been explored \cite{vogt2020subwavelength,wang2019chip}. For example, implementing voltage actuated thermal tuning and the THz magneto-optical effect in InSb allowed for the recent realisation of integrated broadly tunable filters and isolators, respectively \cite{wang2019voltage,yuan2021chip}. While significantly more research is required, these rapid advances highlight the vast potential of silicon THz microresonator devices in the THz frequency range.

Pivotal to the success of silicon THz microresonators is the low material absorption of silicon, as the intrinsic losses of the substrate ultimately limit the Q-factor. The extraordinary low material absorption of high resistivity float zone grown silicon (HRFZ-Si) in the THz domain has been known for a long time \cite{xie2021review,dai2004terahertz,li2008dielectric}; in fact, HRFZ-Si is considered lossless for most applications. However, because of the diminutive absorption it is challenging to precisely measure the material absorption with conventional methods like THz time-domain spectroscopy. This is highlighted by the extensive range of results reported in the literature \cite{xie2021review, dai2004terahertz,li2008dielectric} and references within. Moreover, while small variations of the silicon material absorption might be insignificant for many applications, for silicon THz microresonator devices the material absorption profoundly affects the Q-factor and thus the device performance.

In this work, we systematically investigate the Q-factor of HRFZ-Si THz disc microresonators for a range of resistivities and determine the material absorption of silicon as a function of its resistivity. Our results highlight how critical resistivity is to achieve high-quality silicon THz microresonator devices, and allow a straightforward estimate of the achievable Q-factor based on the substrate's resistivity. In general, the presented methodology can be applied to precisely determine the material absorption of low loss materials in the THz domain.  

%%%%%%%%%%%%%%%%%%%%%%%%%%%%%%%%%%%%%%%%%%
\section*{Methods}
We characterise the Q-factor and resistivity of six HRFZ-Si disc microresonators fabricated from three wafers (two disc microresonators per wafer) at 0.5\,THz and 0.6\,THz, and the experimental results are supported with finite-element simulations as well as an analytical model. The experiments performed at 0.5\,THz and 0.6\,THz serve as a proof-of-concept demonstration; the presented methodology is applicable to the entire THz domain.

\begin{figure}[htbp]
\centering\includegraphics[width=13cm]{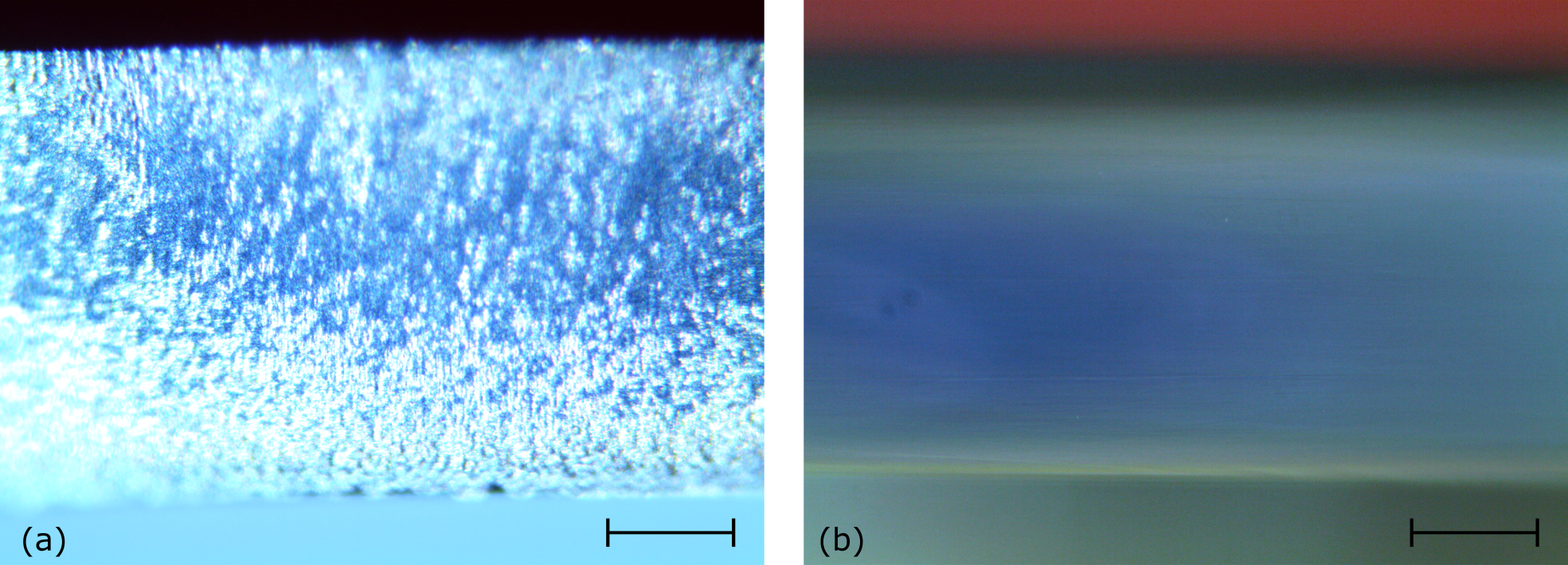}
\caption{(a) 362\,$\upmu$m thick disc after the laser micromachining without polishing. (b) The same disc after several iterative polishing steps with diamond slurry. Please note these images are not taken at the same position on the disc's rim. Both images are taking with an optical microscope with a 20x magnification objective; the scale bar shows 100\,$\upmu$m.}
\label{fig:1}
\end{figure}

Starting with 2 inch to 4 inch diameter HRFZ-Si wafers, the disc microresonators are fabricated using femto-second laser micromachining with 220\,fs pulses at 800\,nm with a repetition rate of 1\,kHz and 20\,mW power. Subsequently, the disc's rim is polished with a fine diamond slurry to remove imperfections caused by the laser micromachining. The diamond slurry polishing leads to slightly rounded edges of the discs; however, this and any radial symmetric features do not deteriorate the intrinsic Q-factors of the resonators. Moreover, to validate the reproducibility of the fabrication process, we characterise two discs from each wafer. Figure \ref{fig:1} (a) and (b) show a typical disc rim before and after the polishing step, respectively. The polishing is necessary to avoid surface scattering losses which would reduce the Q-factor, albeit not completely suppress the resonances. Please note that only the disc's rim is polished as the laser micromachining does not introduce imperfections to the top and bottom surfaces of the disc.

\begin{figure}[t]
\centering\includegraphics[width=12.9cm]{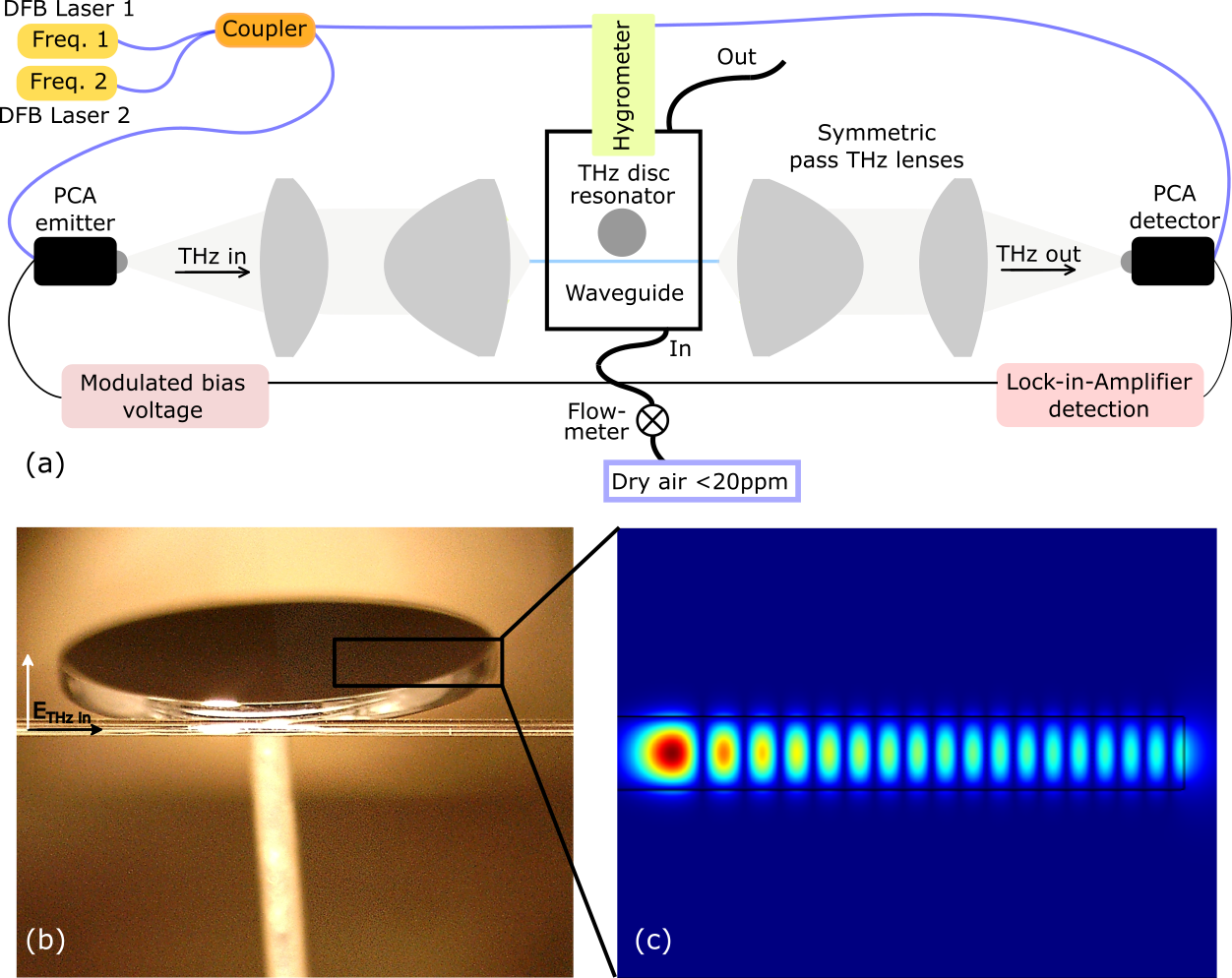}
\caption{(a) Schematic of the experimental setup to study the Q-factor of the investigated THz disc microresonators. The frequency-domain THz spectrometer with photo-conductive antennas (PCAs) is a TeraScan 1550nm from Toptica Photonics \cite{deninger20152}. The high numerical aperture symmetric pass polymer lenses are specifically designed to coupled to the 200\,$\upmu$m diameter air-silica step index waveguide \cite{vogt2018ultra}. (b) Photograph of a 395\,$\upmu$m thick silicon disc in the setup next to the coupling waveguide; the discs are glued to a thin rod ($\approx$500\,$\upmu$m diameter). The black arrow indicates the direction of propagation of the THz field while the white arrow indicates the polarisation. (c) A simulated typical electric field distribution of a higher order radial mode in a 6\,mm diameter 395\,$\upmu$m thick THz disc microresonator at 0.6\,THz, like the one shown in (b). Because the mode is confined to the outer 1.5\,mm of the disc microresonator, only 2\,mm from the rim of the disc is shown, as indicated with the black box in (b). The polarisation is parallel to the disc's rotational axis.}
\label{fig:2}
\end{figure}

The Q-factor of the THz disc microresonators are characterised using coherent THz frequency-domain spectroscopy, and data analysis based on Hilbert transform \cite{vogt2018ultra, vogt2019coherent}. A schematic of the experimental setup is shown in Fig. \ref{fig:2} (a). At the heart of the setup is the single-mode sub-wavelength air-silica step-index waveguide (200\,$\upmu$m diameter) used to evanescently couple the linearly polarised THz radiation to the disc microresonators. The position of the microresonator relative to the waveguide is controlled with a 2D computerised translation stage. The typical coupling distance between the waveguide and disc microresonator depending on the Q-factor and resonance frequency is about 100\,$\upmu$m to 150\,$\upmu$m. In order to avoid absorption from water vapour, the microresonator is placed inside a gas cell with less than 20 part-per-million water vapour as closely monitored with a commercial high-end hygrometer. A photograph of a microresonator close to the waveguide is shown in Fig. \ref{fig:2} (b). The optical surface quality of the disc's rim is clearly visible. 

The intrinsic Q-factor $Q_i$ (Q-factor of the uncoupled resonator) is extracted by fitting an analytical model to the investigated resonances. This process is repeated for six measurements with slightly different coupling positions i.e. loaded Q-factors $Q_l$, and the average of the extracted intrinsic Q-factor from those measurements is given in the manuscript. Whereby the loaded Q-factor $Q_l$ is given by $Q_l = {({Q_i}^{-1}+{Q_c}^{-1})}^{-1}$, with $Q_c$ the coupling Q-factor \cite{gorodetsky1999optical}. The detailed description of the THz frequency domain spectrometer and the data analysis based on the Hilbert transform is described in great detail elsewhere \cite{vogt2019coherent}.      

The resistivity of the discs is measured using a commercial four-point probe setup with tungsten carbide tips to penetrate the native silica layer of silicon. The tips are carefully placed around the centre of the disc, and the finite diameter and thickness of the disc are accounted for in the calculations \cite{smits1958measurement}. Resistivity measurements with positive and negative polarity are performed to ensure ohmic contact with the sample, and the given resistivities are the averages of the two readings. Interestingly, the resistivity measurements reveal large variations even on one wafer as discussed in detail below. For example, a HRFZ-Si wafer specified with a resisitivity >10\,k$\Omega$cm has resistivities of 12\,k$\Omega$cm and 18\,k$\Omega$cm on different positions on the wafer.

The Q-factor measurements are supported with finite element simulations using COMSOL Multiphysics\textsuperscript{\textregistered} software \cite{COMSOL}. In particular, we model the fabricated discs using the measured diameter and thickness and adjust the material absorption until the simulated Q-factor matches the measurements. By doing so, we can link the measured Q-factor and resistivity to the material absorption of silicon; this provides a supplementary method to the analytical model discussed below. The rounded edges from the polishing are not considered in the model; however, as mentioned above, the rounded edges do not impact the Q-factor. Please note that the simulations are Eigenfrequency simulations of the rotational symmetric disc microrseonators in two dimensions and do not contain the coupling waveguide. Figure \ref{fig:2} (c) shows a typical mode profile of a higher-order radial mode experimentally excited in the HRFZ-Si THz disc microresonators. As nearly all of the energy of the different modes is confined within the discs, it is irrelevant which modes exactly are experimentally excited, as variations of the Q-factor for different modes are insignificant compared to the experimentally observed uncertainty.

The finite element simulations also guide the design of the disc microresonators. In particular, two considerations determine the diameter of the discs. On the one hand, the diameter has to be large enough so that the radiation losses are significantly smaller than the losses due to the material absorption. On the other hand, the diameter should be as small as possible for a large free-spectral range to avoid difficulties with an over-population of modes in a narrow frequency range. The intrinsic Q-factor is given as $Q_i ={({Q_m}^{-1}+{Q_r}^{-1})}^{-1}$, with $Q_m$ the Q-factor determined from the material absorption and $Q_r$ the radiation Quality factor. As long as $Q_r$ is significantly larger than $Q_m$, the intrinsic Q-factor is determined by the material absorption which is essential for the presented method. Based on the simulations, a good balance is achieved with 6\,mm diameter HRFZ-Si THz disc microresonators with a minimum disc thickness of 200\,$\upmu$m. For example, for a 6\,mm diameter, and 395\,$\upmu$m thick disc radiation losses are neglectable with a $Q_r$ of $6.4\times{10}^{8}$, which is significantly larger then the maximum anticipated $Q_m$ of about $1\times{10}^{5}$. The disc thicknesses used in this work are 362\,$\upmu$m, 395\,$\upmu$m, and 523\,$\upmu$m - meeting the requirements of neglectable radiation losses. %Please note that the various thicknesses are included in the finite element modelling.   

%\end{verbatim}

\begin{figure}[t]
\centering\includegraphics[width=13cm]{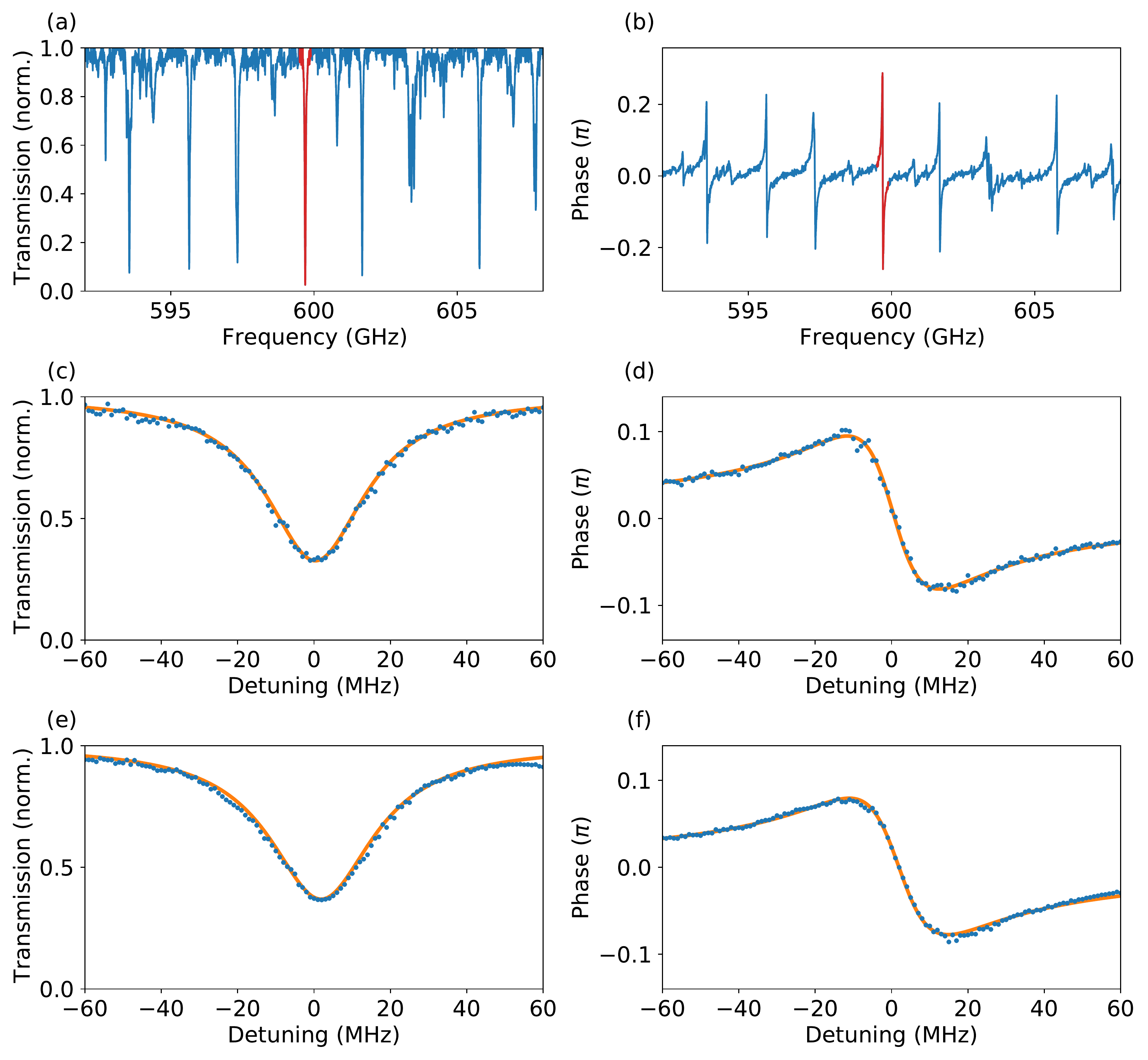}
\caption{(a) Transmission of the air-silica step-index waveguide coupled to a 395\,$\upmu$m thick disc microresonator in the frequency range from 592\,GHz to 608\,GHz, normalised to the waveguide transmission without microresonator. (b) The corresponding phase profile to (a). (c) High resolution scan of the resonance at 599\,GHz [highlighted in red in (a) and (b)]. (d) Shows the corresponding phase profile to (c). Please note that the normalised transmission and phase profile shown in (c) and (d), respectively, are recorded independently of (a) and (b) at a different coupling position. (e) and (f) show the normalised transmission and phrase profile of a resonance of the same THz disc microresonator at 509\,GHz. The lower Q-factor compared to the resonance at 599\,GHz is nicely visible, due to the comparable coupling strengths. The frequency step size in sub-figures (c) to (f) is 1\,MHz.}
\label{fig:3}
\end{figure}

%%%%%%%%%%%%%%%%%%%%%%%%%%%%%%%%%%%%%%%%%%
\section*{Results and Discussion}
A typical normalised transmission and phase profile of the air-silica step-index waveguide coupled to one of the 395\,$\upmu$m thick disc microresonators is shown in Fig. \ref{fig:3} (a) and (b) in the frequency range from 592\,GHz to 608\,GHz, respectively; a multitude of modes are excited with various coupling strengths due to different phase matching and evanescent field overlap \cite{vogt2018ultra}. The highlighted mode (in red) at 599\,GHz is used to extract the intrinsic Q-factor of this disc microresonator, and a high resolution zoom-in is shown in Fig. \ref{fig:3} (c) and (d). As described above, the intrinsic Q-factor is extracted by fitting an analytical model (solid orange line), which agrees very well with the measurements \cite{gorodetsky1999optical}.

\begin{figure}[t]
\centering\includegraphics[width=13cm]{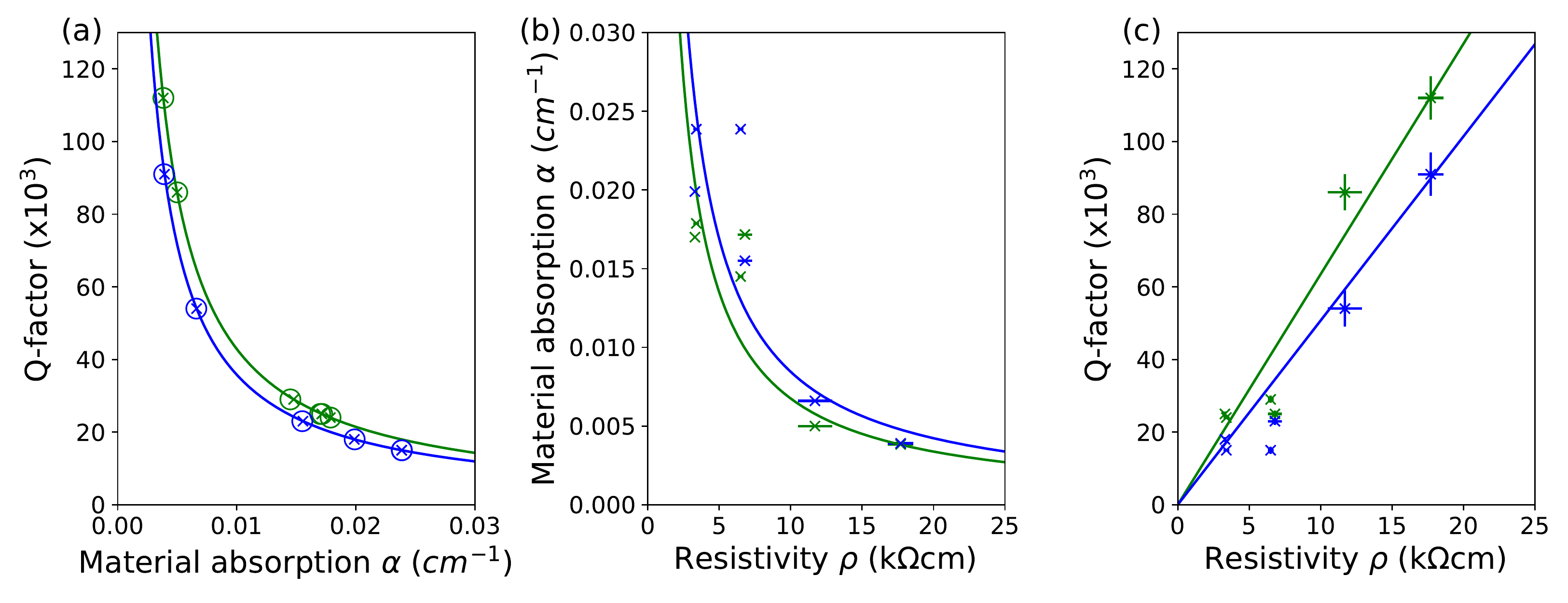}
\caption{(a) Q-factor of the THz HRFZ-Si disc microresonators as a function of the silicon material absorption at 0.5\,THz (blue crosses) and 0.6\,THz (green crosses). The corresponding analytical model and the finite-element simulations are shown with solid lines and circles, respectively. Please note that two data points for each frequency are very close together and therefore only five data points per frequency are visible at the shown scale. (b) Silicon material absorption as a function of the measured resistivity [0.5\,THz (blue) and 0.6\,THz (green)]. The functional dependence of the fitted solid lines is: $\alpha= b/\rho$, with $b=84.5\,\Omega$ at 0.5\,THz and $b=67.5\,\Omega$ at 0.6\,THz (c) Q-factor as a function of resistivity at 0.5\,THz (blue) and 0.6\,THz (green). The solid lines are plotted with Eq. \ref{eq:1} but using the fit from Fig. 4 (b) for the equation.}
\label{fig:4}
\end{figure}

The measured Q-factors at 0.5\,THz (blue crosses) and 0.6\,THz (green crosses) for all discs are summarised in Fig. \ref{fig:4} (a) as a function of the material absorption. The material absorption is calculated from the measured Q-factor based on an analytical model \cite{gorodetsky1999optical}:  

%The Q-factor is inversely proportional to the material absorption, as can be seen from the fits (solid blue and green lines). This is in great agreement with the analytical model (solid black line) : 

\begin{equation}
Q=\frac{2\pi {n}_{s}f}{\mathrm{c} \alpha},
\label{eq:1}
\end{equation}

with ${n}_{s}$ the substrate refractive index of 3.416 \cite{vogt2018thermal}, c the speed of light and $\alpha$ the material absorption. The Q-factor is inversely proportional to the material absorption, and the corresponding curves are plotted in solid blue and green lines for 0.5\,THz and 0.6\,THz, respectively. We confirmed the validity of the analytical model with finite-element simulations, where we adjust the material absorption until the simulated Q-factor matches the experimental observation for the particular disc microresonator. The analytical model and the simulations [blue and green circles in Fig. \ref{fig:4} (a)] are almost in perfect agreement. However, the analytical model is more convenient to use and provides a better understanding.   

Next, Fig. \ref{fig:4} (b) depicts the material absorption as a function of the measured resistivity $\rho$. As previously suggested, the silicon material absorption is inversely proportional to the resistivity because the absorption is predominantly caused by free-charge carriers in silicon \cite{dai2004terahertz}. The experimental results are fitted with $\alpha= b/\rho$, with $b$ the only fitting parameter, with $b=84.5\,\Omega$ at 0.5\,THz and $b=67.5\,\Omega$ at 0.6\,THz. The corresponding fits are shown with solid blue and green lines in Fig. \ref{fig:4} (b). Generally, both frequencies at 0.5\,THz (blue crosses) and 0.6\,THz (green crosses) follow the same trend, albeit slightly higher material absorption at 0.5\,THz. This is in agreement with previously reported results \cite{vogt2018ultra}. Moreover, the material absorption shown in Fig. \ref{fig:4} (b) is in agreement with a previously reported upper limit of 0.1\,${cm}^{-1}$ in the frequency range from 0.2\,THz to 1\,THz \cite{dai2004terahertz}. A direct comparison of the material absorption is unfortunately not possible as commonly used techniques like THz time-domain spectroscopy can only measure upper limits for the minute material absorption of HRFZ-Si.      

Finally, Fig. \ref{fig:4} (c) shows the measured Q-factors as a function of the resistivity measured with the four-point probe setup. Because of the inverse proportionality of the material absorption with resistivity, the Q-factor of the THz disc microresonators is linear with resistivity (see Eq. \ref{eq:1}). The solid blue and green lines in Fig. \ref{fig:4} (c) are plotted based on the analytical model and the fitted material absorption $\alpha$ as a function of the resistivity $\rho$ with fitting parameter $b$.

Importantly, disc microresonators with similar resistivities have almost matching Q-factors, confirming the reproducibility of the fabrication process. Interestingly, however, the 362\,$\upmu$m thick discs, while cut from the same wafer, have resistivities of 12\,k$\Omega$cm and 18\,k$\Omega$cm, which is reflected in the measured Q-factors of 86k and 112k, respectively. The inconsistent resistivities across one wafer are most likely related to minute amount of imperfections in the wafer itself at or after the fabrication. This shows, that individual resistivity measurements of commercially available wafers with a specified resistivity of $>$10\,k$\Omega$cm are necessary to ensure consistent and high-end device performance.

%%%%%%%%%%%%%%%%%%%%%%%%%%%%%%%%%%%%%%%%%%

%%%%%%%%%%%%%%%%%%%%%%%%%%%%%%%%%%%%%%%%%%
\section*{Conclusions}
The presented results highlight the importance of the silicon substrate's resistivity on the intrinsic Q-factor of THz microresonators -  the building blocks for next-generation THz devices. Consistent high-end performance requires rigorous testing of the silicon substrate's resistivity, as the slightest variations - even within one wafer - can lead to significant Q-factor degradation. In particular, our results show that Q-factors of more than 100k can be achieved by rigorous resistivity measurements of the silicon substrate. 

Moreover, the presented methodology is also suitable to measure the material absorption of low-loss materials in the THz domain, which can be challenging using conventional THz spectroscopy. To the best of our knowledge, this is the first time this technique has been applied in the THz domain to precisely measure the material absorption of HRFZ-Si as a function of the substrate's resistivity.  

%%%%%%%%%%%%%%%%%%%%%%%%%%%%%%%%%%%%%%%%%%

\section*{Conflicts of interest}
The authors declare no conflict of interest.

%%%%%%%%%%%%%%%%%%%%%%%%%%%%%%%%%%%%%%%%%%

\bibliography{ref}

%%%%%%%%%%%%%%%%%%%%%%%%%%%%%%%%%%%%%%%%%%

%%%%%%%%%%%%%%%%%%%%%%%%%%%%%%%%%%%%%%%%%%

\end{document}